\documentstyle[prc,aps,manuscript,epsfig]{revtex}
\tightenlines
\begin{document}
\draft
\title{Memory effects on descent from nuclear fission barrier}
\author{V.M. Kolomietz$^{1,2)}$, S.V. Radionov$^{1)}$ and S. Shlomo$^{2)}$}
\address{$^{1)}$Institute for Nuclear Research, Kiev 03028, Ukraine}
\address{$^{2)}$Cyclotron Institute, Texas A\&M University, College Station, 
Texas 77843, USA}
\maketitle

\begin{abstract}
Non-Markovian transport equations for nuclear large amplitude motion are
derived from the collisional kinetic equation. The memory effects are caused
by the Fermi surface distortions and depend on the relaxation time. It is
shown that the nuclear collective motion and the nuclear fission are
influenced strongly by the memory effects at the relaxation time $\tau \geq
5\cdot 10^{-23}{\rm s}$. In particular, the descent of the nucleus from the
fission barrier is accompanied by characteristic shape oscillations. The
eigenfrequency and the damping of the shape oscillations depend on the
contribution of the memory integral in the equations of motion. The shape
oscillations disappear at the short relaxation time regime at $\tau \to 0$,
which corresponds to the usual Markovian motion in the presence of friction
forces. We show that the elastic forces produced by the memory integral lead
to a significant delay for the descent of the nucleus from the barrier.
Numerical calculations for the nucleus $^{236}$U shows that due to the
memory effect the saddle-to-scission time grows by a factor of about 3 with
respect to the corresponding saddle-to-scission time obtained in liquid drop
model calculations with friction forces.
\end{abstract}

\bigskip



\vskip1cm

\section{Introduction}

The dynamics of a nucleus undergoing fission can be studied in terms of only
a few collective variables like nuclear shape parameters \cite{hamy}. Such
kind of approach is usually associated with the liquid drop model (LDM) and
its extensions and is acceptable for a slow collective motion where the fast
intrinsic degrees of freedom exert forces on the collective variables
leading to a Markovian transport equation. An essential assumption is that
the LDM provides a good approximation for a smooth part, $\widetilde{E}_{%
{\rm pot}},$ of the collective potential energy, $E_{{\rm pot}},$ and can be
then used for the quantum calculations of \ $E_{{\rm pot}}$ within
Strutinsky's shell correction method \cite{fanny}, obtaining $E_{{\rm pot}}=%
\widetilde{E}_{{\rm pot}}+\delta {\cal U}$, where $\delta {\cal U}$\ is the
shell correction. On the other hand, it is well known that the LDM is not
able to describe\ some strongly collective nuclear excitations such as the
isoscalar giant multipole resonances. It is because the LDM ignores the
important features of the nucleus as a Fermi liquid. The collective motion
of the nuclear Fermi liquid is accompanied by the dynamical distortion of
the Fermi-surface \cite{ak} and the smooth energy $\widetilde{E}_{{\rm pot}}$%
\ is subsidized by an additional contribution, $\widetilde{E}_{{\rm pot,F}},$
which is caused by the {\it dynamic} Fermi-surface distortion effect and is
absent in the standard LDM \cite{nisi,kiko}. We point out that the energy
$\widetilde{E}_{{\rm pot,F}}$ is a smooth quantity\ (in the sense of the
shell correction method) and\ it can not be recovered by taking into
consideration the quantum shell corrections to the adiabatic (static)
potential energy deformation. This situation becomes more clear in the limit
of the infinite Fermi system with $A\rightarrow \infty $, where $A$ is the
number of particles. The shell correction $\delta {\cal U}$\ disappears at $%
A\rightarrow \infty .$ In this case, the adiabatic collective energy
$E_{{\rm pot}} = \widetilde{E}_{{\rm pot}},$ caused by a change 
of the particle density $\rho$ with respect to its equilibrium value 
$\rho _{{\rm eq}},$ determines the first sound velocity $c_{1}=\sqrt{K/9m},$ 
where $K$ is the incompressibility coefficient given by 
\[
K=\frac{\partial ^{2}(E_{{\rm pot}}/A)}{\partial \rho ^{2}}\rho ^{2}|_{{\rm %
eq}}\approx \frac{\partial ^{2}(\widetilde{E}_{{\rm pot}}/A)}{\partial \rho
^{2}}\rho ^{2}|_{{\rm eq}}.
\]
However, it is well known \cite{lipi,Peth} that, in a cold Fermi liquid, the
first sound velocity $c_{1}$ is relevant only in the limit of strong
interaction, i.e., at $|F_{0}|\gg 1,$ where $F_{0}$ is the Landau parameter
in the quasiparticle scattering amplitude. For the normal nuclear matter we
have rather the value of $F_{0}\sim 0$, which can be derived from the Skyrme
forces \cite{vabr}. The corresponding sound velocity (zero sound velocity) $%
c_{0}$ exceeds the velocity $c_{1}$ by a factor of about $\sqrt{3}$ at $%
F_{0}\sim 0$\ \cite{Peth}. This difficulty is overcome if the additional
contribution $\widetilde{E}_{{\rm pot,F}}$\ to the potential energy $%
\widetilde{E}_{{\rm pot}}$ is taken into account \cite{KoKo99}. Thus, the
smooth energy $\widetilde{E}_{{\rm pot,F}},$ caused by the Fermi surface
distortion effect,\ is a necessary ingredient of the dynamics of the nuclear
Fermi liquid. It is absent in the adiabatic deformation energy $\widetilde{E}%
_{{\rm pot}}$ derived by the traditional LDM. 

The equations of motion for the nuclear Fermi liquid can be derived from the
collisional kinetic equation \cite{ak}. In general, the corresponding
equations of motion are non-Markovian \cite{GrSc88,abay}. The memory effects
appear here due to the Fermi-surface distortion and depend on the relaxation
time \cite{ayik,KoPl92}. The Markovian dynamic is achieved in two limiting
cases only: (i) short relaxation time limit which corresponds to the first
sound propagation in infinite Fermi liquid. In fact, this limit is realized
by the nuclear LDM, (ii) infinite relaxation time limit which corresponds to
the zero sound propagation with a strong renormalization of the sound
velocity and the deformation energy with respect to the ones in the LDM. The
non-Markovian-Langevin equations of motion for macroscopic collective
variables were earlier derived in \cite{aysu} and used for the small
amplitude dynamics \cite{kiko,boab} and for some aspects of \ the induced
nuclear fission and\ the fission rate problem \cite{BoSu93}.

The main purpose of the present paper is to apply the non-Markovian dynamics
to the descent of the nucleus from the fission barrier. Starting from the
collisional Landau-Vlasov kinetic equation, we suggest a new proof of the
non-Markovian equations of motion for the nuclear shape variables which
establishes a direct connection between the memory effects and with the dynamic
distortion of the Fermi surface. In contrast to Ref. \cite{BoSu93}, we do
not take into consideration the random forces and only concentrate on the
formation of \ both the conservative and the friction forces behind the
saddle point to clarify the effects of the memory integral. In this
aspect, our approach represents an extension of the traditional LDM theory
of the nuclear fission \cite{hamy,swiateck,dama,sckn} to the case of the
Fermi liquid and takes into account the important features of the dynamic
Fermi surface distortion which are ignored in the LDM.

The plan of the paper is as follows. In Sec. II we obtain the Euler-like
equation of motion for the displacement field. This equation contains the
memory dependent pressure tensor. Assuming that the nucleus is an
incompressible and irrotational fluid and using the boundary conditions for
the velocity field, we reduce the local Euler-like equation to the
non-Markovian equations of motion for the shape variables. The transport
coefficients and the memory kernel are derived through the solution to the
Neumann problem for the potential of the velocity field. In Sec. III we
study the dependence of the memory effects on the relaxation time for both
the small amplitude motion near the saddle point and for the descent of the
nucleus from the barrier to the scission point. Summary and conclusions are
given in Sec. IV.\ 

\bigskip

\section{Non-Markovian dynamics of nuclear Fermi liquid drop}

To derive the equation of motion for the shape variables, we will start from
the collisional kinetic equation for the phase-space distribution function $%
f\equiv f({\bf r,p;}t)$ in the following general form 
\begin{equation}
{\frac{\partial }{\partial t}}f+{\frac{{\bf p}}{m}}\cdot {\bf \nabla }_{r}f-%
{\bf \nabla }_{r}U\cdot {\bf \nabla }_{p}f=I[f],  \label{1}
\end{equation}
where $U\equiv U({\bf r,p;}t)$ is the selfconsistent mean field and $I[f]$ is
the collision integral. The momentum distribution is distorted during the time
evolution of the system and the distribution function takes the form 
\begin{equation}
f({\bf r,p;}t)=f_{{\rm sph}}({\bf r,p;}t)+\sum_{l\geq 1}\delta f_{l}({\bf %
r,p;}t),  \label{f}
\end{equation}
where $f_{{\rm sph}}({\bf r,p;}t)$ describes the spherical distribution in
momentum space and $l$ is the multipolarity of the Fermi-surface distortion.
We point out that the time dependent Thomas-Fermi (TDTF) approximation and
the corresponding nuclear LDM are obtained from Eq. (\ref{1})  if one takes
the distribution function $f({\bf r,p;}t)$ in the following restricted form $%
f_{{\rm TF}}({\bf r,p;}t)=f_{{\rm sph}}({\bf r,p;}t)+\delta f_{l=1}({\bf r,p;%
}t)$ instead of Eq. (\ref{f}), see Ref. \cite{kota}. Below we will extend
the TDTF approximation taking into account the dynamic Fermi surface
distortion up to multipolarity $l=2$ \cite{nisi,kiko,hoec}. We will also
assume that the collective motion is accompanied by a small deviation of the
momentum distribution from the spherical symmetry, i.e., even in the case of
large amplitude motion the main contribution to the distribution function $f(%
{\bf r,p;}t)$ is given the Thomas-Fermi term $f_{{\rm TF}}({\bf r,p;}t)$ \ and
the additional term $\delta f_{l=2}({\bf r,p;}t)$\ provides only small
corrections. The lowest orders $l=0$ and $1$ (which are not necessary small)
\ of the Fermi-surface distortion do not contribute to the collision
integral because of the conservation laws \cite{ak} and the linearized
collision integral with respect to small perturbation $\delta f_{l=2}({\bf %
r,p;}t),$ is given by 
\begin{equation}
I[f]=-\frac{\delta f_{l=2}}{\tau },  \label{I}
\end{equation}
where $\tau $ is the relaxation time.

Evaluating the first three moments of Eq. (\ref{1}) in ${\bf p}$-space, we
can derive a closed set of equations for the following moments of the
distribution function, namely, local particle density $\rho $, velocity
field $u_{\nu }$ and pressure tensor $P_{\nu \mu }$, in the form (for
details, see Refs. \cite{kota,book}) 
\begin{equation}
{\frac{\partial }{\partial t}\rho }=-{\nabla }_{\nu }(\rho u_{\nu }),
\label{e1}
\end{equation}
\begin{equation}
m\rho {\frac{\partial }{\partial t}}u_{\nu }+m\rho \ (u_{\mu }\nabla _{\mu
})\ u_{\nu }+\nabla _{\nu }{\cal P}+\rho \nabla _{\nu }{\frac{\delta {\cal %
\epsilon }_{{\rm pot}}}{\delta \rho }}=-\nabla _{\mu }P_{\nu \mu }^{\prime },
\label{e2}
\end{equation}
\begin{equation}
{\frac{\partial }{\partial t}}P_{\nu \mu }^{\prime }+{\cal P\ }{\frac{%
\partial }{\partial t}}\Lambda _{\nu \mu }=-\frac{1}{\tau }P_{\nu \mu
}^{\prime },  \label{e3}
\end{equation}
where ${\cal P\equiv P}({\bf r,}t)$ is the isotropic part of the pressure
tensor 
\begin{equation}
{\cal P}({\bf r,}t)={\frac{1}{3m}}\int {\frac{d{\bf p}}{(2\pi \hbar )^{3}}}%
p^{2}f_{{\rm sph}}({\bf r,p;}t),  \label{p}
\end{equation}
$P_{\nu \mu }^{\prime }=P_{\nu \mu }^{\prime }({\bf r,}t)$ is the deviation
of the pressure tensor from its isotropic part, ${\cal P}({\bf r,}t),$\ due
to the Fermi surface distortion 
\begin{equation}
P_{\nu \mu }^{\prime }({\bf r,}t)={\frac{1}{m}}\int {\frac{d{\bf p}}{(2\pi
\hbar )^{3}}}(p_{\nu }-mu_{\nu })(p_{\mu }-mu_{\mu })\delta f_{l=2}({\bf r,p;%
}t),  \label{p1}
\end{equation}
${\cal \epsilon }_{{\rm pot}}$ is the potential energy density related to
the selfconsistent mean field $U$ as $U=\delta {\cal \epsilon }_{{\rm pot}%
}/\delta \rho .$ The tensor $\Lambda _{\nu \mu }$\ in Eq. (\ref{e3}) is
given by 
\begin{equation}
\Lambda _{\nu \mu }=\nabla _{\nu }\chi _{\mu }+\nabla _{\mu }\chi _{\nu }-{%
\frac{2}{3}}\delta _{\nu \mu }\nabla _{\lambda }\chi _{\lambda },
\label{lambda}
\end{equation}
where $\chi _{\nu }\equiv \chi _{\nu }({\bf r,}t)$ is the displacement field
related to the velocity field as $u_{\nu }\equiv $ $u_{\nu }({\bf r,}%
t)=\partial \chi _{\nu }({\bf r,}t)/\partial t.$ From Eq. (\ref{e3}) we find
the pressure tensor $P_{\nu \mu }^{\prime }({\bf r,}t)$ in the following
form 
\begin{equation}
P_{\nu \mu }^{\prime }({\bf r,}t)=P_{\nu \mu }^{\prime }({\bf r,}%
t_{0})-\int_{t_{0}}^{t}dt^{\prime }\exp (\frac{t^{\prime }-t}{\tau })\ {\cal %
P}({\bf r,}t^{\prime })\ \frac{\partial }{\partial t^{\prime }}\Lambda _{\nu
\mu }({\bf r,}t^{\prime }).  \label{p11}
\end{equation}
The tensor $P_{\nu \mu }^{\prime }({\bf r,}t_{0})$\ is determined by the
initial conditions. In the case of the quadrupole distortion of the Fermi
surface, the tensor $P_{\nu \mu }^{\prime }({\bf r,}t_{0})$ is derived by
the initial displacement field $\chi _{\nu }.$

Assuming that the nucleus is\ an incompressible and irrotational fluid with
a sharp surface in ${\bf r}$-space, we will reduce the local equation of
motion (\ref{e2}) to the equations for the variables $q=%
\{q_{1},q_{2},.....q_{N}\}$ that specify the shape of the nucleus. The
continuity equation (\ref{e1}) has to be complemented by the boundary
condition on the moving nuclear surface $S$. Below we will assume that the
axially symmetric shape of the nucleus is defined by rotation of the profile
function $\rho =Y(z,\{q_{i}(t)\})$ around the $z$-axis in the cylindrical
co-ordinates $\rho ,z,\varphi $ \cite{dasi76,ivko}. The velocity of the
nuclear surface is then given by \cite{ivko} 
\begin{equation}
u_{S}=\sum_{i=1}^{N}\bar{u}_{i}\dot{q}_{i},  \label{us}
\end{equation}
where 
\begin{equation}
\bar{u}_{i}=({{\partial Y}/{\partial q_{i}}})/\Lambda ,\ \ \ \Lambda =\sqrt{%
1+({{\partial Y}/{\partial z}})^{2}}.  \label{ui1}
\end{equation}
The potential of the velocity field takes the form 
\begin{equation}
\phi =\sum_{i=1}^{N}{}_{i}\overline{\phi }\ \dot{q}_{i},  \label{phi2}
\end{equation}
where the potential field $\overline{\phi }_{i}\equiv \overline{\phi }_{i}(%
{\bf r,}q)$ is determined by the equations of the following Neumann problem 
\cite{ivko} 
\begin{equation}
\nabla ^{2}\stackrel{\_}{\phi }_{i}=0\,,\,\,\ \ \ ({\bf n}\nabla \stackrel{\_%
}{\phi }_{i})_{S}={\frac{1}{\Lambda }}{\frac{{\partial Y}}{{\partial q_{i}}}}%
,  \label{neum}
\end{equation}
where ${\bf n}$ is the unit vector which is normal to the nuclear surface.

Using Eqs. (\ref{e2}) and (\ref{p11}) with $u_{\nu }=\nabla _{\nu }\phi ,$
multiplying Eq. (\ref{e2}) by $\nabla _{\mu }\overline{\phi }_{i}$ and
integrating over ${\bf r}$, one obtains 
\begin{equation}
\sum_{j=1}^{N}[B_{ij}(q)\ddot{q}_{j}+\sum_{k=1}^{N}\frac{\partial B_{ij}}{%
\partial q_{k}}\ \stackrel{\cdot }{q_{j}}\stackrel{\cdot }{q_{k}}%
+\int_{t_{0}}^{t}dt^{\prime }\exp (\frac{t^{\prime }-t}{\tau })\kappa
_{ij}(t,t^{\prime })\ \stackrel{\cdot }{q_{j}}(t^{\prime })]=-\frac{\partial
E_{{\rm pot}}(q)}{\partial q_{i}}.  \label{main}
\end{equation}
Here $B_{ij}(q)$ is the inertia tensor 
\begin{equation}
B_{ij}(q)=m\rho _{0}\oint ds\bar{u}_{i}\ \overline{\phi }_{j},  \label{bij}
\end{equation}
where $\rho _{0}$ is the nuclear bulk density. The adiabatic collective
potential energy, $E_{{\rm pot}}(q),$ does not contain the contribution from
the Fermi-surface distortion effect and is given by 
\begin{equation}
E_{{\rm pot}}(q)=\int d{\bf r\ (}{\cal \epsilon }_{{\rm kin}}({\bf r},q)+%
{\cal \epsilon }_{{\rm pot}}({\bf r},q)),  \label{epot1}
\end{equation}
where ${\cal \epsilon }_{{\rm kin}}({\bf r},q)$\ is the kinetic energy of
the internal motion of nucleons. The memory kernel $\kappa
_{i,j}(t,t^{\prime })$\ in Eq. (\ref{main}) is given by 
\begin{equation}
\kappa _{ij}(t,t^{\prime })=2\ \int d{\bf r\ }{\cal P}({\bf r,}q(t^{\prime
}))\ (\nabla _{\nu }\nabla _{\mu }\stackrel{\_}{\phi }_{i}({\bf r,}q(t)))\
(\nabla _{\nu }\nabla _{\mu }\stackrel{\_}{\phi }_{j}({\bf r,}q(t^{\prime
}))).  \label{kappa1}
\end{equation}

In Eq. (\ref{main}), we have omitted the contribution from the initial
distortion of the Fermi surface caused by the pressure tensor $P_{\nu \mu
}^{\prime }({\bf r,}t_{0}).$ The contribution from $P_{\nu \mu }^{\prime }(%
{\bf r,}t_{0})$ reflects the fact that the initial displacement field $\chi
_{\mu }({\bf r,}t_{0})$ is switched on suddenly at $t=t_{0}$. The adiabatic
force $-\ \partial E_{{\rm pot}}(q)/\partial q$ in Eq. (\ref{main})\ obtains
then the additional contribution\ at $t=t_{0}$\ due to the initial\
distortion of the Fermi surface. However the corresponding force is absent
if the initial displacement field $\chi _{\mu }({\bf r,}t_{0})$ is obtained
as a result of the previous evolution of the system at $t<t_{0}$. Below we
will consider the descent of the nucleus from the fission barrier, i.e.
assume the presence of the evolution of the system at $t<t_{0},$ and omit
the contribution from the pressure tensor $P_{\nu \mu }^{\prime }({\bf r,}%
t_{0})$.

The displacement field $\chi _{\nu }({\bf r,}q)$ and the potential field $%
\overline{\phi }_{i}\equiv \overline{\phi }_{i}({\bf r,}q)$ are determined
by a solution to the Neumann problem (\ref{neum}). The displacement field $%
\chi _{\nu }({\bf r,}q)$ can be also obtained using the Werner-Wheeler
method \cite{dasi76}. In the cylindrical co-ordinates $\rho ,z,\varphi ,$
the components of velocity field $u_{z}$ and $u_{\rho }$ in $z$ and $\rho $
directions are then approximated as \cite{dasi76} 
\begin{equation}
u_{z}=\sum_{i}{\cal A}_{i}(z,q)\dot{q}_{i}\,\,\,\,\,,\,\,\,\,u_{\rho }={%
\frac{\rho }{{Y(z,q)}}}\sum_{i}{\cal B}_{i}(z,q)\dot{q}_{i}.
\end{equation}
The two unknown coefficients ${\cal A}_{i}(z,q)$ and ${\cal B}_{i}(z,q)$ are
related to each other by means of continuity equation as 
\begin{equation}
{\cal B}_{i}(z,q)=-{\frac{1}{2}}Y(z,q){\frac{{\partial {\cal A}_{i}(z,q)}}{{%
\partial z}}}.
\end{equation}
Requiring then that the normal velocity of the fluid on the surface should
coincide with the normal velocity of the surface one can express the
coefficient ${\cal A}_{i}(z,q)$ in terms of the profile function $Y(z,q)$ as 
\begin{equation}
{\cal A}_{i}(z,q)=Y^{-2}(z,q){\frac{\partial }{{\partial q_{i}}}}%
\int_{z}^{z_{{\rm \max }}}dz^{\prime }Y^{2}(z^{\prime },q).
\end{equation}
We point out that in the case of irrotational flow the Werner-Wheeler method
leads to a velocity field potential of quadrupole type \cite{ivko} 
\begin{equation}
\phi ({\bf r,}q)=\frac{1}{4q}\ (2\ z^{2}-x^{2}-y^{2}).  \label{phi3}
\end{equation}
A spheroidal figure presents the simplest example which is consistent with
the velocity field potential (\ref{phi3}). In this case $q=q(t)$ is the
elongation of the figure in units of the radius $R_{0}=r_{0}A^{1/3}$ of the
nucleus and the equation of motion (\ref{main}) takes the following form

\begin{equation}
B(q)\stackrel{\cdot \cdot }{q}+\frac{\partial B(q)}{\partial q}\stackrel{%
\cdot }{q}^{2}=-\frac{\partial E_{{\rm pot}}(q)}{\partial q}%
-\int_{t_{0}}^{t}dt^{\prime }\exp (\frac{t^{\prime }-t}{\tau })\kappa
(t,t^{\prime })\ \stackrel{\cdot }{q}(t^{\prime }).  \label{eq2}
\end{equation}
Here, the mass parameter $B(q)$ and the memory kernel $\kappa (t,t^{\prime
}) $\ are given by

\begin{equation}
B(q)=\frac{1}{5}\ A\ m\ R_{0}^{2}\ \left( 1+\frac{1}{2q^{3}}\right) \ \ {\rm %
and}\ \ \ \kappa (t,t^{\prime })=\frac{\kappa _{0}}{q(t)\ q(t^{\prime })},
\label{bkappa}
\end{equation}
where $\kappa _{0}=(4/5\ m)\ \pi \ \rho _{0}\ p_{F}^{2}\ R_{0}^{3}$ and $%
p_{F}$ is the Fermi momentum.

\section{Numerical calculations and discussion}

Let us start from the one-dimension case and apply Eq. (\ref{eq2}) to the
large amplitude motion from the barrier point {\rm B }to the ''scission''
point{\rm \ C} in Fig. 1. Following the Kramers model \cite{kram}, we will
approximate the potential energy $E_{{\rm pot}}(q)$ by an upright oscillator 
$(1/2)C_{LDM}(q-q_{0})^{2}$ with $q_{0}=1$ and an inverted oscillator $%
E_{f}-(1/2)\widetilde{C}_{LDM}(q-q_{f})^{2}$ which are joined smoothly\ as
shown in Fig. 1 (see also Ref. \cite{nisi2}). Let us consider, first of all,
a small amplitude change, $\Delta q$, of the shape variable, $q,$ near both
the ground state at $q\sim q_{0}=1$ with $\Delta q=q-q_{0}$ and at the
saddle point at $q\sim q_{f}$ with $\Delta q=q-q_{f}$. Linearizing Eq. (\ref
{eq2}), we will rewrite it as 
\begin{equation}
\widetilde{B}\ \frac{\partial ^{2}}{\partial t^{2}}\Delta q\ =-\ k\ \Delta q-%
\widetilde{\kappa }\int_{t_{0}}^{t}dt^{\prime }\exp (\frac{t^{\prime }-t}{%
\tau })\ \frac{\partial }{\partial t^{\prime }}\Delta q(t^{\prime }),
\label{eq4}
\end{equation}
where $\widetilde{B}=B_{0}\equiv B(q=1),\ k=C_{LDM},$ $\widetilde{\kappa }%
=\kappa _{0}$ if $q\sim q_{0}$ and $\widetilde{B}=B_{f}\equiv B(q=q_{f}),\
k=-\widetilde{C}_{LDM},$ $\widetilde{\kappa }=\kappa _{f}=\kappa
_{0}/q_{f}^{2}$ if $q\sim q_{f}.$ Differentiating Eq. (\ref{eq4}) over time,
we will look for the solution to Eq. (\ref{eq4}) in the form 
\begin{equation}
\Delta q=\sum_{i=1}^{3}C_{i}\exp (\lambda _{i}t).  \label{q2}
\end{equation}
Here the coefficients $C_{i}$ are derived by the initial conditions. The
eigenvalues $\lambda _{i}$\ are obtained as a solution to the following
secular equation 
\begin{equation}
(\lambda ^{2}+\frac{k}{\widetilde{B}})(\lambda +\frac{1}{\tau })+\frac{%
\widetilde{\kappa }}{\widetilde{B}}\lambda =0.  \label{disp4}
\end{equation}
In the case of the zero relaxation time limit, $\tau \rightarrow 0,$ one
obtains from Eq. (\ref{disp4}) a non-damped motion with $\lambda =\pm \sqrt{%
k/\widetilde{B}}$, i.e., the time evolution is derived by the LDM stiffness
coefficients $C_{LDM}$ or $\widetilde{C}_{LDM}$. In the opposite case of
rare collisions, $\tau \rightarrow \infty ,$ the solution to Eq. (\ref{disp4}%
) leads to a non-damped motion with $\lambda =\pm \sqrt{(k+\widetilde{\kappa 
})/\widetilde{B}}.$ In contrast to the previous case, the additional
contribution, $\widetilde{\kappa },$ appears at the stiffness coefficient $k+%
\widetilde{\kappa }$ because of the Fermi surface distortion effect. In the
case of the nuclear Fermi liquid one has $\widetilde{\kappa }\gg |k|$ \cite
{nisi,kiko}. This fact is important for the description of the nuclear
isoscalar giant resonances \cite{nisi,kiko}. Considering a motion near the
ground state at $q\sim q_{0}=1$ with $\ k=C_{LDM},$ one obtains from Eq. \ (%
\ref{disp4}) the quadrupole eigenvibrations with the eigenenergy \ 
\begin{equation}
\hbar \omega _{2^{+}}=\hbar \sqrt{\frac{C_{LDM}+\kappa _{0}}{B_{0}}}\approx
\hbar \sqrt{\frac{4\epsilon _{F}}{mR_{0}^{2}}}\approx 64.5\ A^{-1/3}\ {\rm %
MeV},  \label{en1}
\end{equation}
where $\epsilon _{F}=(9\pi )^{2/3}\hbar ^{2}/8mr_{0}^{2}=34.73\ {\rm MeV}$
is the Fermi energy and we adopt $r_{0}=1.18$ ${\rm fm}$. The result (\ref
{en1}) coincides with the analogous one obtained earlier\ by Nix and Sierk 
\cite{nisi} and agrees with the experimental value of the energy of the
isoscalar quadrupole resonance $\hbar \omega _{2^{+}}^{{\rm \exp }}\approx
63\cdot A^{-1/3}\ {\rm MeV}$.

As can be seen from Eq. (\ref{disp4}), the motion is damped for the non-zero
and finite relaxation time $\tau .$ In the case of small amplitude motion
near the ground state, $q\sim q_{0},$ the solution to Eq. (\ref{eq4}) at $%
t-t_{0}\gg \tau $ takes the form of eigenvibrations with $\Delta q(t)\sim
\exp (i\omega t)$, where the eigenfrequency $\omega $\ is derived by \cite
{kiko} 
\begin{equation}
\omega ^{2}B_{0}=C_{LDM}+C^{\prime }(\omega )-i\omega \gamma (\omega ),
\label{disp5}
\end{equation}
where $B_{0}=B(q=1),$ the additional stiffness coefficient, $C^{\prime
}(\omega ),$ appears due to the Fermi surface distortion effect 
\begin{equation}
C^{\prime }(\omega )=\kappa _{0}%
\mathop{\rm Im}%
\left( \frac{\omega \tau }{1-i\omega \tau }\right)   \label{c11}
\end{equation}
and the friction coefficient $\gamma (\omega )$ is given by 
\begin{equation}
\gamma (\omega )=\kappa _{0}%
\mathop{\rm Re}%
\left( \frac{\tau }{1-i\omega \tau }\right) =(4/m)\ \pi \ R_{0}^{3}\ \eta
_{0}%
\mathop{\rm Re}%
\left( \frac{1}{1-i\omega \tau }\right) .  \label{gam1}
\end{equation}
Here, $\eta _{0}=(1/5)\ \rho _{0}\ p_{F}^{2}\ \tau ,$ is the classical
viscosity coefficient \cite{ak}.

Let us consider now the small amplitude motion (starting path for the
development of the instability) near the saddle point, $q\sim q_{f},$ at
finite relaxation time. We have evaluated numerically the value of $\Delta q$
from Eq. (\ref{eq2}) using the secular equation (\ref{disp4}) and the
initial conditions $\Delta q(t_{0})=0,$ $\Delta \stackrel{\cdot }{q}%
(t_{0})=v_{0}$ and $\Delta \stackrel{\cdot \cdot }{q}(t_{0})=0.$ In Fig. 2
we show the result for two values of the relaxation time $\tau =3\cdot
10^{-23}{\rm s}$ and $\tau =4\cdot 10^{-22}{\rm s}$. We have used the
following parameters $q_{f}=1.6,\ A=236$ and $\ \hbar \omega _{f}=\hbar 
\sqrt{|\widetilde{C}_{LDM}|/B_{f}}=1.16$\ ${\rm MeV.}$ The initial velocity $%
v_{0}$\ was derived using the initial kinetic energy $\ E_{{\rm kin,0}%
}=(1/2)B_{f}v_{0}^{2}=1\ {\rm MeV.}$ In the case of the very short
relaxation time, $\tau =3\cdot 10^{-23}{\rm s,}$ the memory effects in Eq. (%
\ref{eq4}) play a minor role only\ and the amplitude of motion is
approximately an exponentially growing function, similar to the case of
Newton motion from the barrier in the presence of the friction forces, see
curve 1 in Fig. 2. The friction coefficient $\gamma $ can be derived here
from Eq. (\ref{eq4}) at $\omega _{F,f}\ \tau \ll 1$ and it is given by $%
\gamma =\gamma _{f}=$ $\kappa _{f}\ \tau =$ $\omega _{F,f}^{2}\ B_{f}\ \tau
\sim \tau ,$ where $\omega _{F,f}=\sqrt{\kappa _{f}/B_{f}}$ is the
characteristic frequency for the eigenvibrations caused by the Fermi surface
distortion effect.\ The behavior of $\Delta q(t)$ is changed dramatically
with an increase of the relaxation time. At large enough relaxation time,
the descent from the barrier is accompanied by the damped oscillations
(curve 2 in Fig. 2). These oscillations are due to the memory integral in
Eq. (\ref{eq4}). The characteristic frequency, $\omega _{R},$ and the
corresponding damping parameter, $\omega _{I},$ can be derived from the
imaginary and real parts of complex conjugated roots of Eq. (\ref{disp4}) as 
$\lambda =-\omega _{I}\pm i$ $\omega _{R}.$ The solution (\ref{q2}) takes
then the form 
\begin{equation}
\Delta q=C_{\zeta }\ e^{\zeta t}+A_{\omega }e^{-\Gamma t/2\hbar }\sin
(Et/\hbar )+B_{\omega }e^{-\Gamma t/2\hbar }\cos (Et/\hbar ),  \label{q4}
\end{equation}
where $\Gamma =2\omega _{I}\hbar $ and $E=\omega _{R}\hbar .$ In Fig. 3 we
show the dependence of the instability growth rate parameter $\zeta ,$\ the
energy of eigenvibrations $E$\ and the damping parameter $\Gamma $\ on the
relaxation time $\tau $.

In the rare collision regime $\omega _{F,f}\ \tau \gg 1,$ the friction
coefficient $\gamma $\ is obtained from Eq. (\ref{eq4}) as $\gamma =\gamma
_{f}=$ $B_{f}\ /\tau \sim 1/\tau .$ We point out that the $\tau $-dependence
of the friction coefficient, $\gamma _{f}\sim 1/\tau ,$ in the rare
collision regime is opposite to the $\tau $-dependence of $\gamma _{f}\sim
\tau $\ in the frequent collision regime. This is a consequence of the
memory effects in the Fermi liquid. Below we will use the following
extrapolation form for the friction coefficient near the fission barrier

\begin{equation}
\gamma _{f}=\omega _{F,f}\ B_{f}\ \frac{\omega _{F,f}\ \tau }{1+(\omega
_{F,f}\ \tau )^{2}}.  \label{gamma2}
\end{equation}

The presence of the memory effects in the equation of motion (\ref{eq4})
changes significantly the trajectory of the nuclear descent from the fission
barrier. The result of the solution of Eq. (\ref{eq4}) for the trajectory $%
\stackrel{\cdot }{q}(q)$ for the large amplitude motion from the saddle
point $q_{f}$ is shown in Fig. 4 (solid line). The dashed line in Fig. 4
shows the trajectory obtained as a solution to the Newton's equation (no
memory effect) 
\begin{equation}
B(q)\stackrel{\cdot \cdot }{q}+\frac{\partial B(q)}{\partial q}\stackrel{%
\cdot }{q}^{2}\ =-\ \frac{\partial E_{{\rm pot}}(q)}{\partial q}-\gamma _{f}%
\stackrel{\cdot }{q},  \label{new2}
\end{equation}
where the friction coefficient $\gamma _{f}$\ was taken from Eq. (\ref
{gamma2}). In both cases we have used the initial conditions with $%
q(t_{0})=q_{f}$, $\stackrel{\cdot }{q}(t_{0})=\sqrt{2E_{{\rm kin,0}}/B_{f}}$
and $E_{{\rm kin,0}}=1\ {\rm MeV}$ and the relaxation time $\tau =4\cdot
10^{-22}{\rm s.}$ As seen from Fig. 4, the memory effect leads to the drift
of $q$ in time which is accompanied by the time oscillations of $q$ along
the trajectory of descent to the ''scission'' point, $q_{{\rm sc}}$. In Fig.
4, the time oscillations of $q$ appear as a spiral-like behavior of the
trajectory $\stackrel{\cdot }{q}(q)$. In both cases,\ the drift from the
barrier is caused by the conservative force $-\ \partial E_{{\rm pot}%
}(q)/\partial q.$ The oscillations appear due to the presence of the
time-reversible elastic force in the memory integral in Eq. (\ref{eq4}), see
also Fig. 2. We point out that\ the memory integral contains the
time-irreversible part also. Due to this fact, the velocity of the system
decreases and the trajectory is shifted to the slope of the fission barrier.
This effect is significantly stronger in the presence of the memory effects
and leads to an essential delay of the descent process with respect to the
analogous result obtained from the Newton's motion of Eq. (\ref{new2}). The
influence of the memory effect on the descent time $t_{{\rm sc}}$ from the
barrier to the ''scission'' point $q_{{\rm sc}}$ is shown in Fig. 5. As seen
from Fig. 5, in the absence of the memory effects (dashed lines), the descent
time $t_{{\rm sc}}$ is about $1\div 3\cdot 10^{-21}{\rm s}$ and, as it
should be, the value of $t_{{\rm sc}}$\ goes to the limit of non-friction
motion for both the frequent collision regime, $\tau \rightarrow 0,$ and the
rare collision regime, $\tau \rightarrow \infty .$ This property of the
descent with no-memory effects\ is the result of the Fermi-liquid
approximation (\ref{gamma2}) for the friction coefficient $\gamma _{f}$ in
Eq. (\ref{new2}). In contrast to this case, the descent time $t_{{\rm sc}}$
evaluated in the presence of the memory effects 
(solid lines) grows monotonously
with the relaxation time $\tau .$ The additional delay of the motion in the
rare collision region (large $\tau )$ is here caused by the contribution of
the elastic force due to the memory integral. The elastic force leads to the
dynamical renormalization of the adiabatic force $-\ \partial E_{{\rm pot}%
}(q)/\partial q$ in Eq. (\ref{eq2}) and acts against the force $-\ \partial
E_{{\rm pot}}(q)/\partial q$.

Let us apply our approach to the case of symmetric nuclear fission described
by Eq. (\ref{main}), assuming the Lorentz parameterization for the profile
function $Y(z)$ in Eq. (\ref{ui1}) in the following form \cite{hamy}, 
\begin{equation}
Y^{2}(z)=(z^{2}-\zeta _{0}^{2})(z^{2}+\zeta _{2}^{2})/Q\,,  \label{shape}
\end{equation}
where the multiplier $Q$ guarantees the volume conservation,
\begin{equation}
Q=-[\zeta _{0}^{3}({\frac{1}{5}}\zeta _{0}^{2}+\zeta _{2}^{2})]/R_{0}^{3}\,.
\end{equation}
Here all quantities of the length dimension are expressed in the $R_{0}$
units. The parameter $\zeta _{0}$ in (\ref{shape}) determines the general
elongation of the figure and $\zeta _{2}$ is related to the radius of the neck.
For $\zeta _{2}=\infty $ the shapes (\ref{shape}) coincide with the
spheroidal ones. At finite $\zeta _{2}$ ($\zeta _{2}>0$ for bound figures)
the neck appears and the value $\zeta _{2}=0$ corresponds to the scission
point after which the figure is divided in the two parts for $\zeta _{2}<0$.
To solve Eq. (\ref{main}) we will rewrite it as a set of two equations.
Namely, 
\begin{equation}
\sum_{j=1}^{2}[B_{ij}(q)\ddot{q}_{j}+\sum_{k=1}^{2}\frac{\partial B_{ij}}{%
\partial q_{k}}\ \stackrel{\cdot }{q_{j}}\stackrel{\cdot }{q_{k}}]=-\frac{%
\partial E_{{\rm pot}}(q)}{\partial q_{i}}+R_{i}(t,q)  \label{main1}
\end{equation}
and 
\begin{equation}
\frac{\partial R_{i}(t,q)}{\partial t}=-\frac{R_{i}(t,q)}{\tau }%
+\sum_{j=1}^{2}\kappa _{ij}(q,q)\dot{q}_{j}\ \ \qquad {\rm at\qquad }%
R_{i}(t=0,q)=0,  \label{main2}
\end{equation}
where $q=\{q_{1},q_{2}\}=\{\zeta _{0},\zeta _{2}\}$ and the terms $\sim \dot{%
q}_{i}\dot{q}_{j}$ were omitted in Eq. (\ref{main2}), as the next order
corrections. The kernel $\kappa _{ij}(q,q)$ is given by 
\begin{equation}
\kappa _{ij}(q,q)=\frac{2}{5}m\rho _{0}v_{F}^{2}\ \int d{\bf r\ }(\nabla
_{\nu }\nabla _{\mu }\stackrel{\_}{\phi }_{i}({\bf r,}q))\ (\nabla _{\nu
}\nabla _{\mu }\stackrel{\_}{\phi }_{j}({\bf r,}q)).  \label{kappa2}
\end{equation}

We have performed numerical calculation for symmetric fission of the nucleus 
$^{236}{\rm U}$. We solved Eqs. (\ref{main1}) and (\ref{main2}) numerically
using the deformation energy $E_{{\rm pot}}(q)$ from Refs. \cite{hamy,myers}%
. The scission line was derived from the condition of the instability of the
nuclear shape with respect to the variations of the neck radius: 
\begin{equation}
\frac{\partial ^{2}E_{{\rm pot}}(q)}{\partial \rho _{{\rm neck}}^{2}}=0
\label{inst}
\end{equation}
where $\rho _{{\rm neck}}=\zeta _{2}/\sqrt{\zeta _{0}({\zeta _{0}^{2}}%
/5+\zeta _{2}^{2})}$ is the neck radius. The equations of motion (\ref{main1}%
) and (\ref{main2}) were solved with the initial conditions corresponding to
the saddle point deformation and the initial kinetic energy $E_{{\rm kin,0}%
}=1~{\rm MeV}$ (initial neck velocity $\dot{\zeta _{2}}=0$). To solve the
Neumann problem (\ref{neum}) for the velocity field potential we have used
the method based on the theory of the potential, see Ref. \cite{ivko}.

In Fig. 6 we show the dependence of the fission trajectory, i.e., the
dependence of the neck parameter $\zeta _{2}$ on the elongation $\zeta _{0},$
for the fissioning nucleus $^{236}{\rm U}$ for two different values of the
relaxation time $\tau $: $\tau =4\cdot 10^{-22}{\rm s}$ (dashed line) and $%
\tau =0$ (dotted line). The scission line (dot-dashed line in Fig. 6) was
obtained as a solution to Eq. (\ref{inst}). We define the scission point as
the intersection point of the fission trajectory with the scission line. As
can be seen from Fig. 6 the memory effect hinders slightly the neck
formation and leads to a more elongated scission configuration. To
illustrate the memory effect on the observable values we have evaluated the
translation kinetic energy of the fission fragments at infinity, $E_{{\rm kin%
}}$, and the prescission Coulomb interaction energy, $E_{{\rm Coul}}$. The
value of $E_{{\rm kin}}$ is the sum of the Coulomb interaction energy at
scission point, $E_{{\rm Coul}}$, and the prescission kinetic energy $E_{%
{\rm kin,ps}}$. Namely,

\begin{equation}
E_{{\rm kin}}{\rm =}E_{{\rm Coul}}+E_{{\rm kin,ps}}{\rm .}  \label{hhh}
\end{equation}
After scission the fission fragments were described in terms of two equal
mass spheroids (see Ref. \cite{swiateck}). We assumed that the distance
between the centers of mass, $d$, of two spheroids is equal to the distance
between the two halves of the fissioning nucleus at the scission point: 
\begin{equation}
d=\frac{5}{4}\zeta _{0}\frac{\zeta _{0}^{2}+3\zeta _{2}^{2}}{\zeta
_{0}^{2}+5\zeta _{2}^{2}}{\rm \mid _{scis}.}  \label{d}
\end{equation}
The corresponding velocity $\dot{d}$ was obtained by the differentiation of
Eq. (\ref{d}) with respect to the time. The elongation, $c$, of both
separated spheroids is defined by the condition: 
\begin{equation}
2c+d=2\zeta {\rm _{0,scis}}  \label{totallen}
\end{equation}
where $\zeta _{0,{\rm scis}}$ is the elongation of the nucleus at the
scission point. The collective parameters $c$ and $d$\ and the velocity $%
\dot{d}$ were then used to evaluate the Coulomb energy $E{\rm _{Coul}}$ (see
Ref. \cite{hamy}) and the prescission kinetic energy $E{\rm _{kin,ps}}$ in
Eq. (\ref{hhh}).

The influence of the memory effects on the fission-fragment kinetic energy, $%
E_{{\rm kin}},$\ and the prescission Coulomb interaction energy, $E_{{\rm %
Coul}},$\ is shown in Fig. 7. As seen from Fig. 7 the memory effects are
neglected at the short relaxation time regime where the memory integral is
transformed into the usual friction force. In\ the case of the Markovian
motion with friction (dashed line), the yield of the potential energy, $%
\Delta E_{{\rm pot}}$, at the scission point is transformed into both the
prescission kinetic energy, $E{\rm _{kin,ps},}$ and the time irreversible
dissipation energy, $E_{{\rm dis}}$, providing $\Delta E_{{\rm pot}}=E{\rm %
_{kin,ps}}$ $+\ E_{{\rm dis}}{\rm .}$ In contrast to this case, the
non-Markovian motion with the memory effects (solid line) produces an
additional time reversible prescission\ energy, $E_{{\rm F,ps}},$ caused by
the distortion of the Fermi surface. In this case, the energy balance reads $%
\Delta E_{{\rm pot}}=E{\rm _{kin,ps}}$ $+\ E_{{\rm dis}}+E_{{\rm F,ps}}.$ We
point out that the two-spheroid parametrization of the fissioning nucleus at 
the scission point given by Eqs. (\ref{d}) and (\ref{totallen}), used in this
work, leads to the prescission Coulomb energy $E_{{\rm Coul}}$ which is about 
5 ${\rm MeV}$ lower (for $^{236}{\rm U}$) than the Coulomb interaction energy
of the scission point shape \cite{dama}. Taking into account this fact and
using the experimental value of the fission-fragment kinetic energy $E_{{\rm %
kin}}^{{\rm \exp }}=168$ ${\rm MeV}$ \cite{dama}, one can see from Fig. 7
that the Markovian motion with friction (dashed line) leads to the
overestimate of the fission-fragment kinetic energy $E_{{\rm kin}}.$ In the
case of the non-Markovian motion with the memory effects (solid line), a
good agreement with the experimental data is obtained at the relaxation time
of about $\tau =8\cdot 10^{-23}{\rm s.}$ A small deviation of the prescission
Coulomb energy $E_{{\rm Coul}}$ obtained at the non-Markovian motion (solid
line in Fig. 7) from the one at the Markovian motion (dashed line in Fig. 7)
is caused by the corresponding deviation of both fission trajectories in
Fig. 6.

In Fig. 8 we illustrate the memory effect on the saddle-to-scission time $t_{%
{\rm sc}}.$ In the case of the non-Markovian motion (solid line), the delay
in the descent of the nucleus from the barrier grows with the relaxation
time $\tau $ (at $\tau \geq 4\cdot 10^{-23}{\rm s}$). This is mainly due to
the hindering action of the elastic force caused by the memory integral. The
saddle-to-scission time increases by a factor of about 2 due to the memory
effect at the ''experimental'' value of the relaxation time $\tau =8\cdot
10^{-23}{\rm s}$ which was derived from the fit of the fission-fragment
kinetic energy $E_{{\rm kin}}$ to the experimental value of $E_{{\rm kin}}^{%
{\rm \exp }}$\ (see above).

\bigskip

\section{Summary and conclusions}

By use of ${\bf p}$-moments techniques, we have reduced the collisional
kinetic equation to the equations of motion for the local values of particle
density, velocity field and pressure tensor. The obtained equations are
closed due to the restriction on the multipolarity $l$ of the Fermi surface
distortion up to $l=2$. To apply our approach to the nuclear large amplitude
motion, we have assumed that the nuclear liquid is incompressible and
irrotational. We have derived the velocity field potential, $\phi ({\bf r,}q)
$, which depends then on the nuclear shape parameters $q(t)$ due to the
boundary condition on the moving nuclear surface. Finally, we have reduced
the problem to a macroscopic equation of motion for the shape parameters $%
q(t).$ Thus we consider a change (not necessary small) of the nuclear shape
which is accompanied by a small quadrupole distortion of the Fermi surface.
The obtained equations of motion for the collective variables $q(t)$
contains the memory integral which is caused by the Fermi-surface distortion
and depends on the relaxation time $\tau $.

The memory effects on the nuclear collective motion disappear in two limits;
of zero relaxation time, $\tau \rightarrow 0,$ and at $\tau \rightarrow
\infty .$ In general case, the memory integral contains the contribution
from both the time reversible elastic force and the dissipative friction
force. The eigenmotion near the ground state (point {\rm A} in Fig. 1) is
influenced by memory effects through the frequency dependency of the
stiffness, $C^{\prime }(\omega ),$ and the friction, $\gamma (\omega ),$
coefficients in the dispersion equation (\ref{disp5}). We point out that the
friction coefficient $\gamma (\omega )$ in our approach (see Eq. (\ref{gam1}%
)) changes its $\tau $-dependency from $\gamma \sim \tau $ in the frequent
collision regime, $\omega _{R}\tau \ll 1,$ to $\gamma \sim 1/\tau $ in the
rare collision regime, $\omega _{R}\tau \gg 1.$ Due to this fact, we have
obtained a correct description of the zero-to-first sound transition in the
nuclear Fermi-liquid \cite{kiko}. In the limit of $\tau \rightarrow \infty ,$
the additional contribution (elastic force) in Eq. (\ref{eq4}) appears due
to the memory integral. The contribution from the elastic force is
significantly stronger than the one caused by the adiabatic force $-\ k\
\Delta q.$ The presence of the elastic force provides a correct $A$%
-dependence of the energy of the isoscalar giant multipole resonances.

We have shown that the development of instability near the fission barrier
(point {\rm B} in Fig. 1) is strongly influenced by the memory effects if
the relaxation time $\tau $\ is large enough. In this case, a drift of the
nucleus from the barrier to the scission point is accompanied by
characteristic shape oscillations (see Figs. 2 and 3) which depend on the
parameter $\widetilde{\kappa }$ of the memory kernel and on the relaxation
time $\tau $. The shape oscillations appear due to the elastic force induced
by the memory integral. The elastic force\ acts against the adiabatic force $%
-\ \partial E_{{\rm pot}}(q)/\partial q$\ and hinders the motion to\ the
scission point {\rm C}. In contrast to the case of the Markovian motion, the
delay in the fission is caused here by the conservative elastic force and
not only by the friction force. Due to this fact, the nucleus loses a part
of the prescission kinetic energy converting it into the potential energy of
the Fermi surface distortion instead of the time-irreversible heating of the
nucleus. As mentioned above, in the nuclear Fermi liquid the friction
coefficient $\gamma $ is a non-monotonic function of the relaxation time $%
\tau $ (see Eqs. (\ref{gam1}) and (\ref{gamma2})) providing the asymptotic
behavior $\gamma \sim \tau $ and $\gamma \sim 1/\tau $ in both limiting
cases of the frequent and rare collisions, respectively. This feature of $%
\gamma $ leads to the non-monotonic behavior of the saddle-to-scission time, 
$t_{{\rm sc}},$ as function of $\tau $ in the case of the Markovian (no
memory) motion with friction, see dashed lines in Figs. 5 and 8. In contrast
to the Markovian motion, the memory effects provide a monotonous dependence
of the saddle-to-scission time on the relaxation time $\tau $ (see solid
lines in Figs. 5 and 8). This is caused by the elastic forces produced by
the memory integral, which lead to the additional hindrance for the descent
from the barrier at large $\tau .$

The memory effects lead to the decrease of the fission-fragment kinetic
energy, $E_{{\rm kin}},$ with respect to the one obtained from the Markovian
motion with friction, see Fig. 7. This is because a significant part of the
potential energy at the scission point is collected as the energy of the
Fermi surface deformation. Note that the decrease of the fission-fragment
kinetic energy due to the memory effects is enhanced in the\ rare collision 
regime (at larger relaxation time) while the effect due to friction decreases.
An additional source for the decrease of the fission-fragment kinetic energy 
is caused by the shift of the scission configuration to that with a larger 
elongation parameter $\zeta _{0},$ in the case of the non-Markovian motion, 
see Fig. 6. Due to this fact, the repulsive Coulomb energy of the fission 
fragments at the scission point decreases with respect to the case of the 
Markovian motion.

\section*{Acknowledgments}

This work was supported in part by the US Department of Energy under grant
\# DOE-FG03-93ER40773. We are grateful for this financial support. One of us
(V.M.K.) thanks the Cyclotron Institute at Texas A\&M University for the
kind hospitality.

\newpage

\begin{figure}[tbp]
\centerline{\epsffile{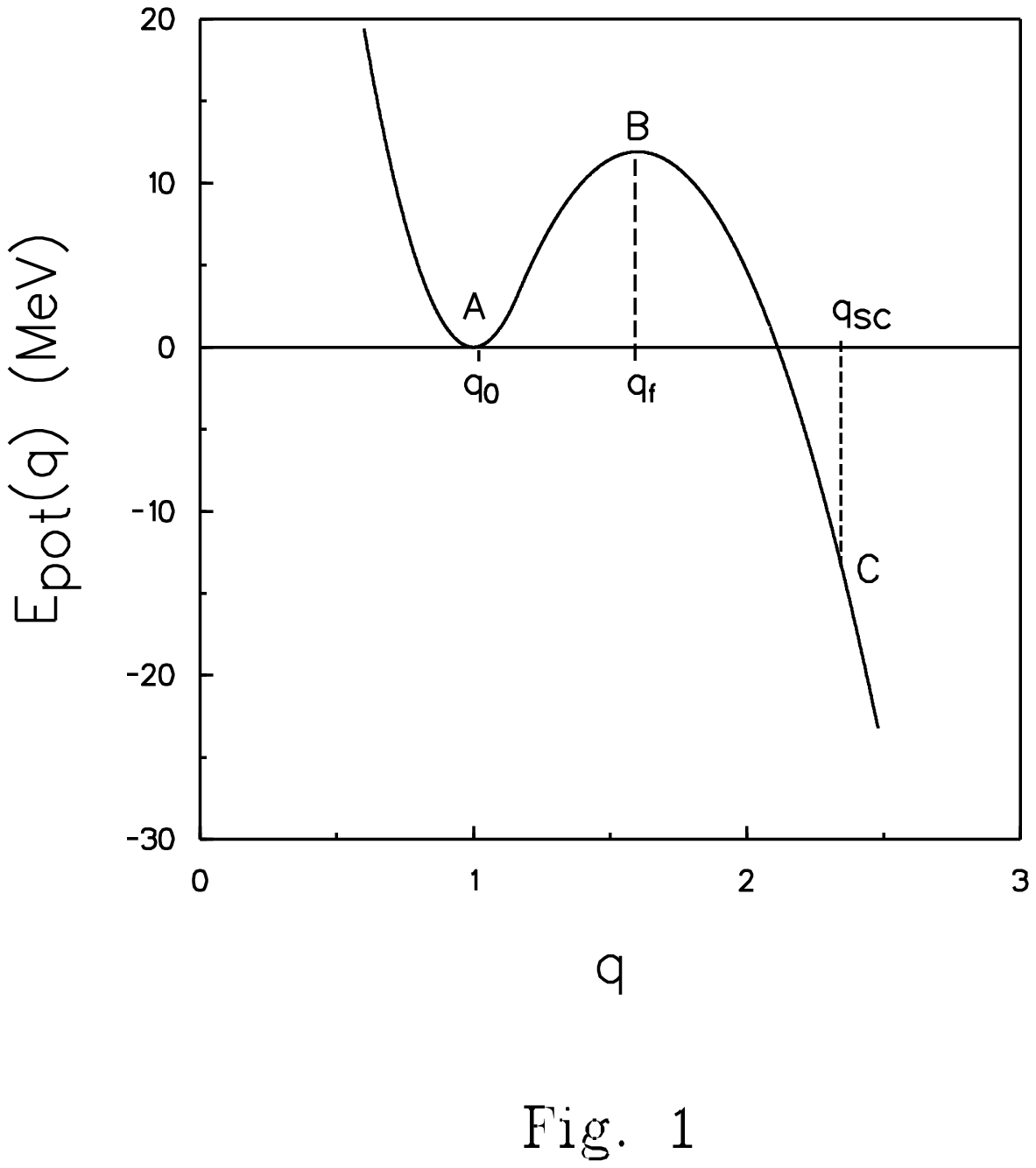}}
\caption{Dependence of the potential energy $E_{{\rm pot}}$ on the shape
parameter $q$.}
\end{figure}

\begin{figure}[tbp]
\centerline{\epsffile{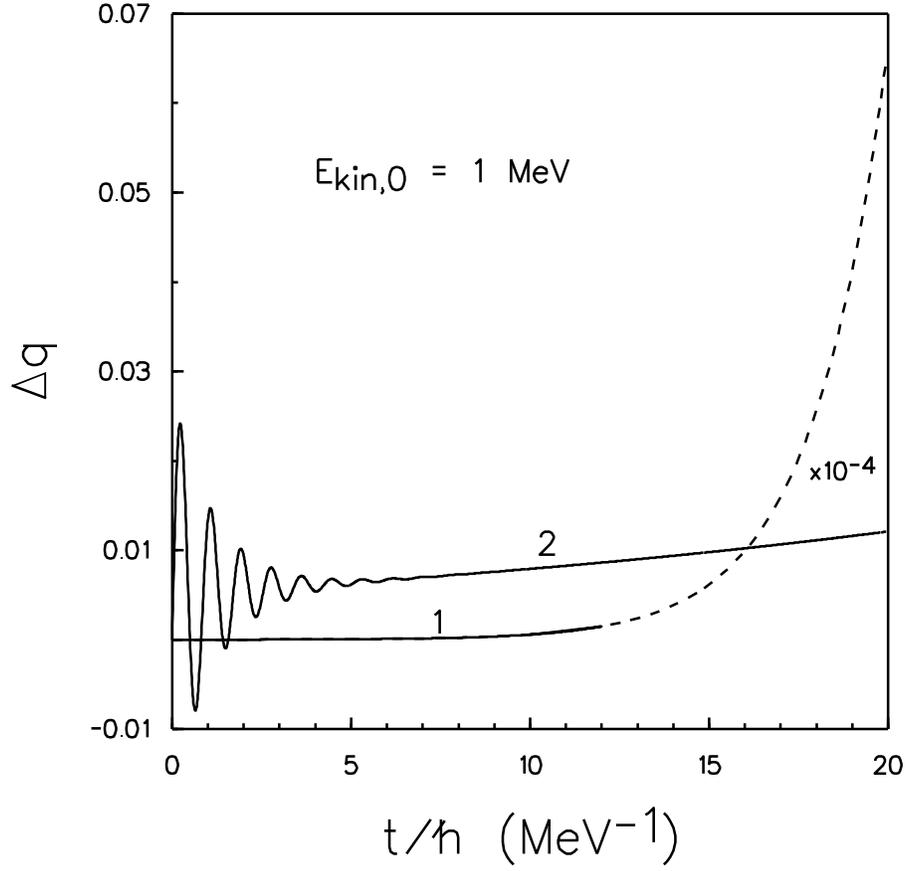}}
\caption{Time variation of the shape parameter $q$ near the saddle point B
(see Fig. 1), for varios values of the relaxation time $\protect\tau$. The
curves 1 and 2 correspond to the values of $\protect\tau = 3\cdot 10^{-23}%
{\rm s}$ $\protect\tau = 4\cdot 10^{-22}{\rm s}$, respectively.}
\end{figure}

\begin{figure}[tbp]
\centerline{\epsffile{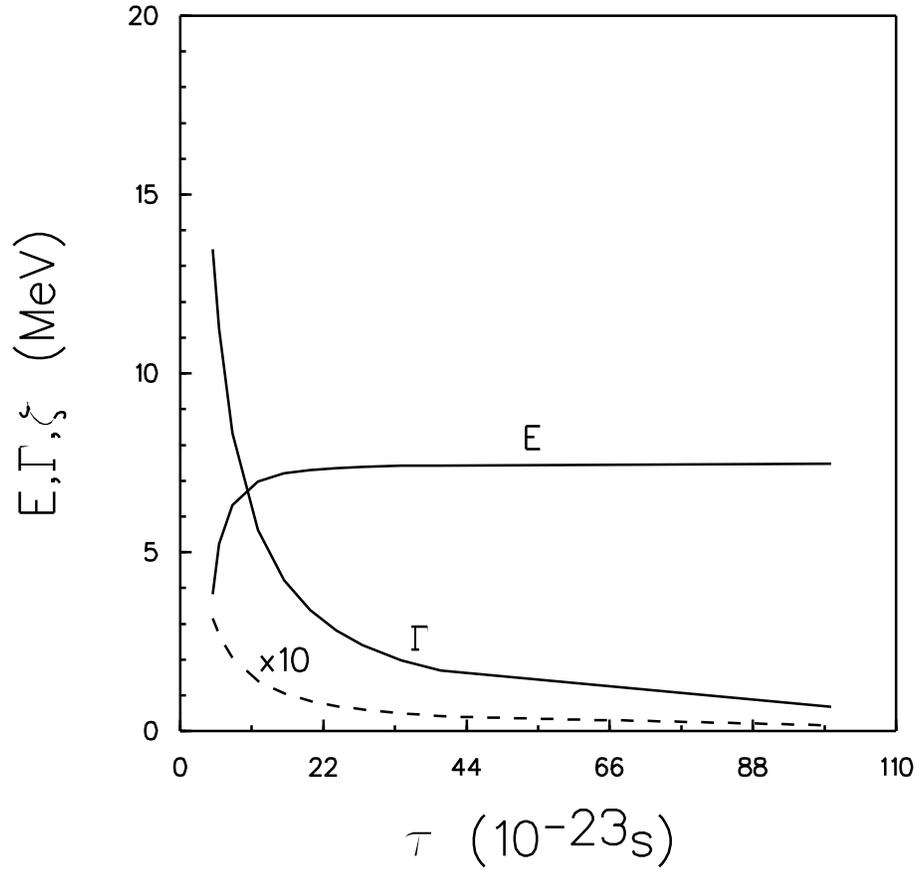}}
\caption{Dependence upon relaxation time $\protect\tau$ of the
characteristic energy $E$ and width $\Gamma$ of oscillations (solid lines)
and the instability growth rate parameter $\protect\zeta$ (dashed line) for
the curve 2 in Fig. 2.}
\end{figure}

\begin{figure}[tbp]
\centerline{\epsffile{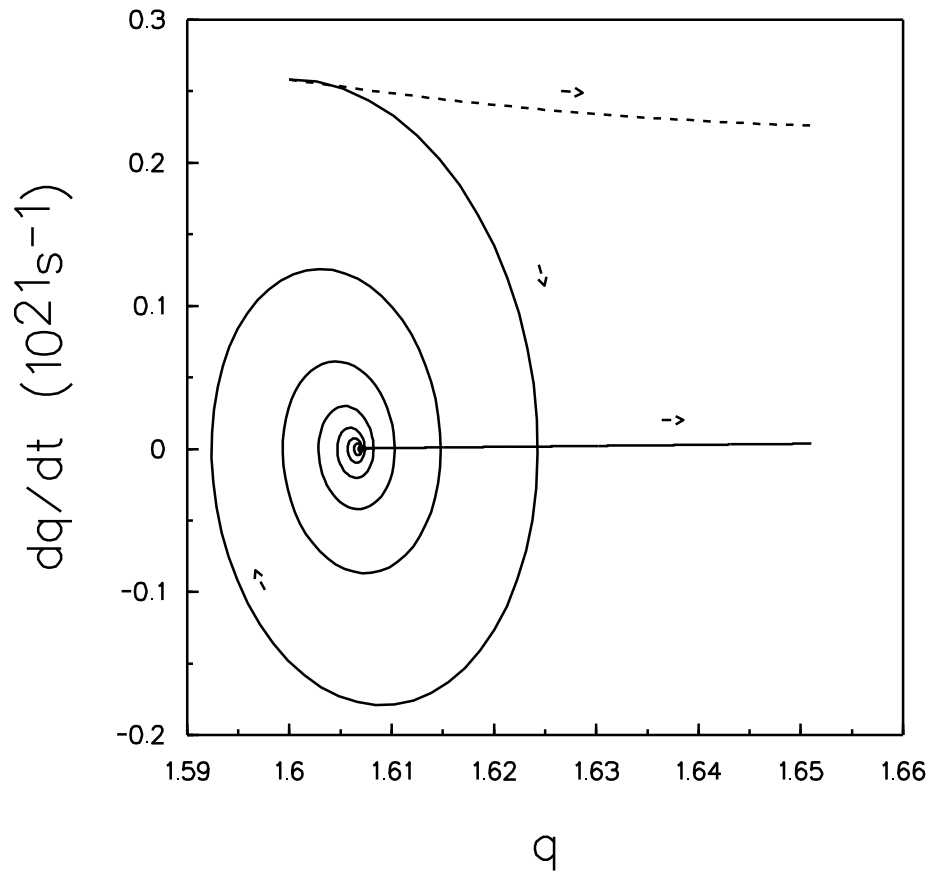}}
\caption{Trajectory (dependence of the collective velocity $dq/dt$ on the
collective coordinate $q$) for the descent from the saddle point B (see Fig.
1). Solid line represents the result of the calculation in presence of the
memory effects and dashed line is for the case of Markovian (no memory)
motion with the friction forces. We have used the relaxation time $\protect%
\tau = 4\cdot 10^{-22}{\rm s}$ and the initial kinetic energy $E_{{\rm kin}%
}=1\,{\rm MeV}$. }
\end{figure}

\begin{figure}[tbp]
\centerline{\epsffile{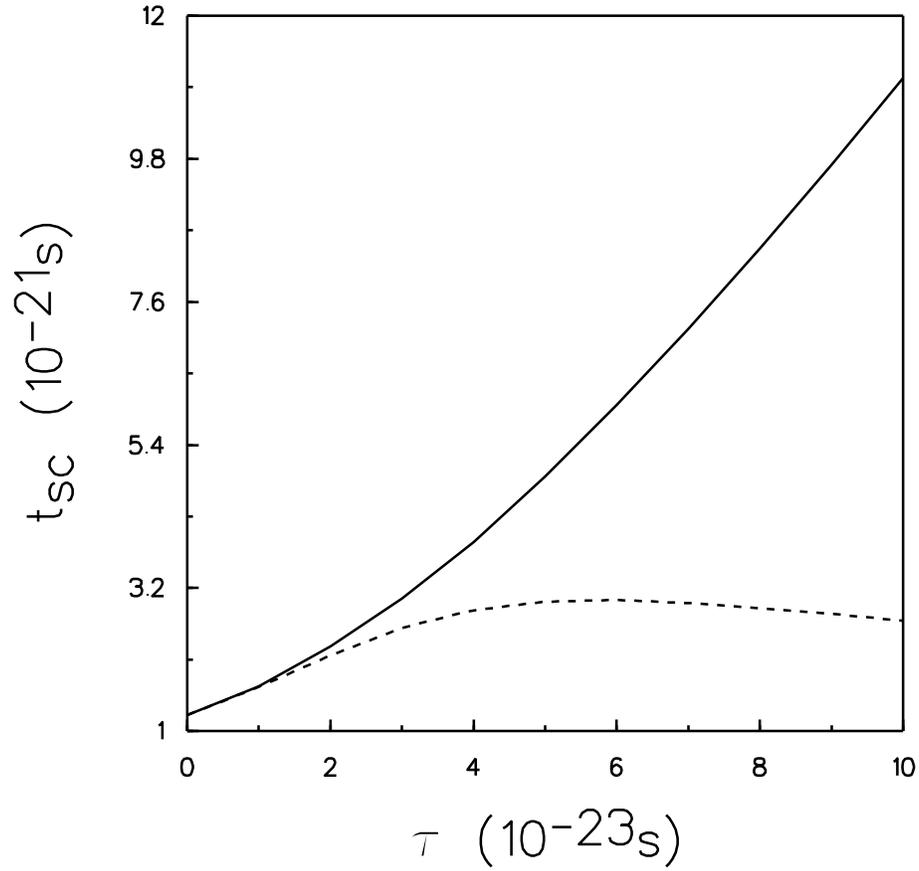}}
\caption{Dependence upon relaxation time $\protect\tau$ of the time, $t_{%
{\rm sc}}$, required to travel a nucleus from the saddle point B to the
"scission" point C (see Fig. 1). Solid line represents the result of the
calculation in presence of the memory effects and dashed line is for the
case of Markovian (no memory) motion with the friction forces. The initial
kinetic energy is $E_{{\rm kin}}=1\,{\rm MeV}$. }
\end{figure}

\begin{figure}[tbp]
\centerline{\epsffile{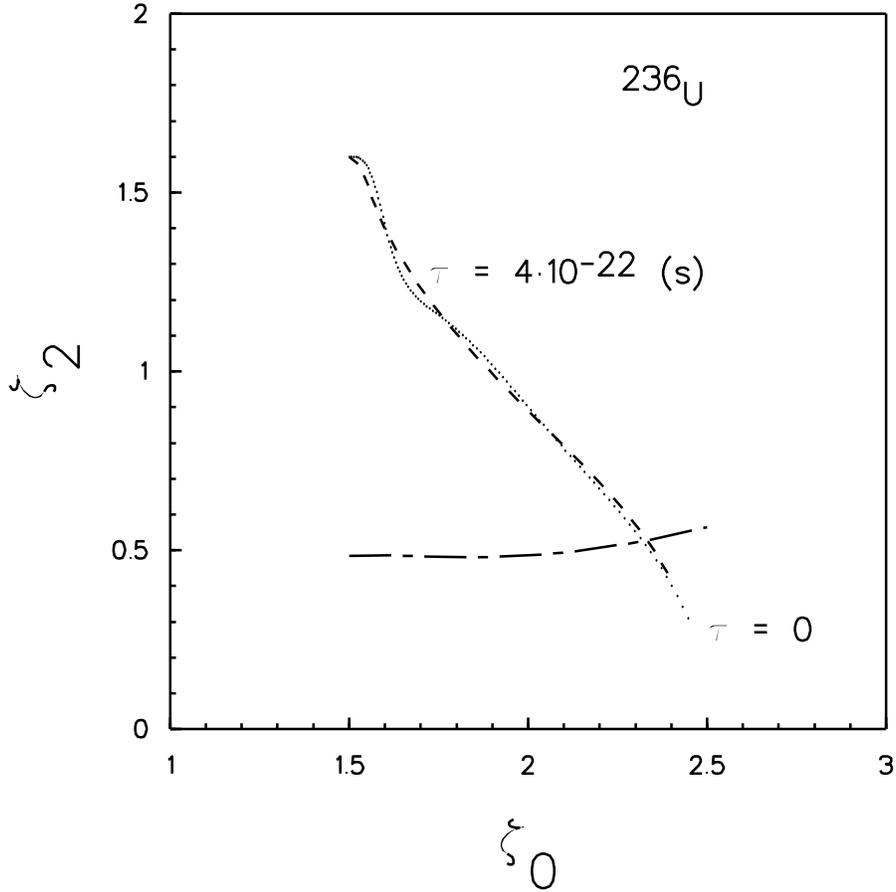}}
\caption{Trajectories of descent from the saddle point of the nucleus $%
^{236}$U in the $\protect\zeta_{0}, \protect\zeta_{2}$ plane. Dashed line
represents the result of the calculation in presence of the memory effects
and dotted line is for the case of Markovian (no memory) motion with the
friction forces. We have used the relaxation time $\protect\tau = 4\cdot
10^{-22}{\rm s}$ and the initial kinetic energy $E_{{\rm kin}}=1\,{\rm MeV}$%
. Dot-dashed line is the scission line derived from the condition (43).}
\end{figure}

\begin{figure}[tbp]
\centerline{\epsffile{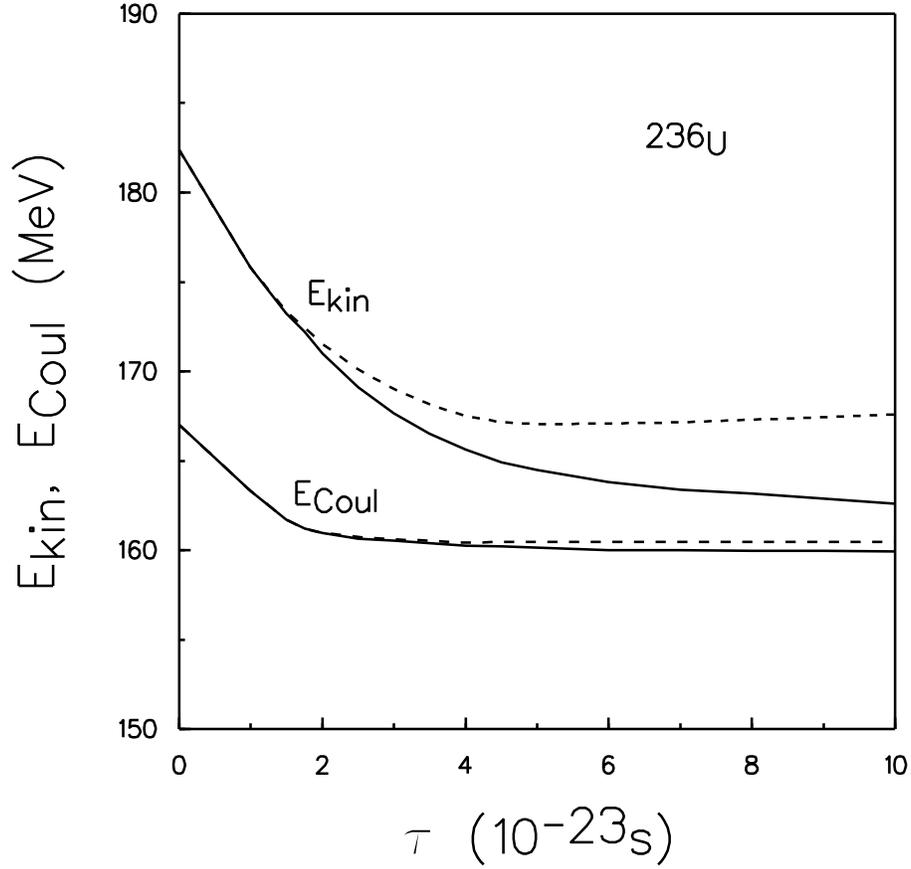}}
\caption{Fission-fragment kinetic energy, $E_{{\rm kin}}$, (curves 1) and
the Coulomb repulsive energy at the scission point, $E_{{\rm Coul}}$,
(curves 2) versus the relaxation time $\protect\tau$ for the nucleus $%
^{236}$U. Solid lines represent the result of the calculation in presence
of the memory effects and dashed lines are for the case of Markovian (no
memory) motion with the friction forces. The initial kinetic energy is $E_{%
{\rm kin, 0}}=1\,{\rm MeV}$.}
\end{figure}

\begin{figure}[tbp]
\centerline{\epsffile{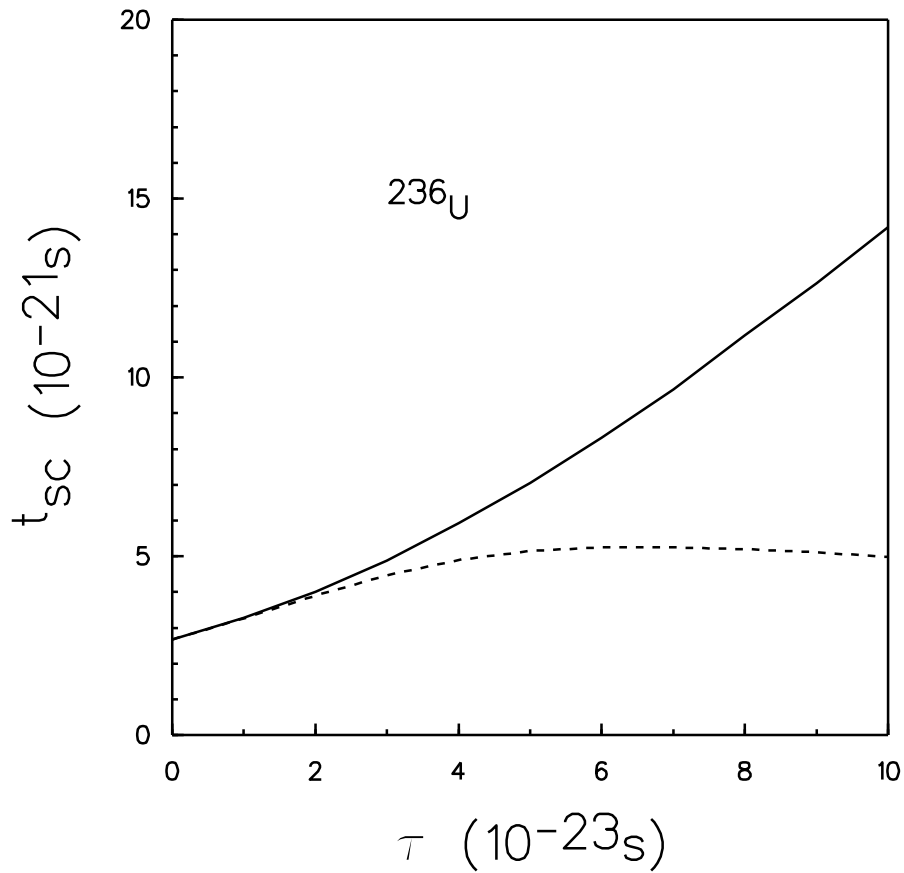}}
\caption{Dependence upon relaxation time $\protect\tau$ of the
saddle-to-fission time, $t_{{\rm sc}}$, for the descent from the barrier in
the case of two-dimension ($\protect\zeta_{0},\protect\zeta_{2}$)
parametrization for the nucleus $^{236}$U. Solid line represents the result
of the calculation in presence of the memory effects and dashed line is for
the case of Markovian (no memory) motion with the friction forces. The
initial kinetic energy is $E_{{\rm kin}}=1\,{\rm MeV}$. }
\end{figure}

\end{document}